\documentclass[letter]{aa}

\usepackage{graphicx}
\usepackage{txfonts}
\usepackage{hyperref}
\usepackage{breakurl}
\begin{document}

   \title{
   A puzzling 2-hour X-ray periodicity in the 1.5-hour orbital period black widow PSR J1311$-$3430
   }


   \author{Andrea De Luca
          \inst{1,2}
          \and
          Martino Marelli
          \inst{1}
          \and
          Sandro Mereghetti
          \inst{1}
          \and
          Ruben Salvaterra
          \inst{1}
          \and 
          Roberto Mignani
          \inst{1}
          \and
          Andrea Belfiore
          \inst{1}
          }

   \institute{INAF--Istituto di Astrofisica Spaziale e Fisica Cosmica di Milano, 
      via A. Corti 12, 20133 Milano, Italy\\
      \email{andrea.deluca@inaf.it}
        \and
             INFN, Sezione di Pavia, via A. Bassi 6, I-27100 Pavia, Italy
             }

   \date{Received --; accepted --}

 
  \abstract
  {Time-domain analysis of an archival {\em XMM-Newton} observation unveiled a very unusual variability pattern in the soft X-ray emission of PSR J1311$-$3430, a black widow millisecond pulsar in a tight binary ($P_B=93.8$ min) with a very low-mass ($M\sim0.01$ $M_{\odot}$) He companion star, known to show flaring emission in the optical and in the X-rays. A series of six pulses with a regular recurrence time of $\sim124$ min is apparent in the 0.2$-$10 keV light curve of the system, also featuring an initial, bright flare and a quiescent phase lasting several hours. The X-ray spectrum does not change when the pulses are seen and is consistent with  a power law with photon index $\Gamma\sim1.6$, also describing the quiescent emission. The peak luminosity of the pulses is of several $10^{32}$ erg s$^{-1}$. Simultaneous observations in the U band with the Optical Monitor onboard XMM and in the g' band from the Las Cumbres Observatory do not show any apparent counterpart of the pulses and only display the well-known orbital modulation of the system. 
  We consider different hypotheses to explain the recurrent pulses: 
  we investigate their possible analogy with other phenomena already observed in this pulsar and in similar systems and we also study possible explanations related to the interaction of the energetic pulsar wind with intrabinary material, but  we found none of these pictures to be convincing. 
  We identify simultaneous X-ray  observations and optical spectroscopy as a possible way to constrain the nature of the phenomenon.
  }

   \keywords{X-rays: stars --
                pulsars: individual: PSR J1311$-$3430 --
                binaries: close
               }
\titlerunning{Puzzling periodicity from PSR J1311$-$3430}
\authorrunning{A. De Luca et al.}
   \maketitle
%

\section{Introduction}

The population of rotation-powered millisecond pulsars (MSP) features a sizeable fraction  of binary systems  (slightly less than 60\%, according to the ATNF pulsar database\footnote{\url{https://www.atnf.csiro.au/research/pulsar/psrcat/}}). Among them, of particular interest are those with very tight orbits ($P_B\ll$1 day) and very low-mass companion stars, classified \citep[see e.g. ][]{roberts13} as ``black widows'' (BW, companion star mass $<0.1\,M_{\odot}$) and ``redbacks'' (RB, companion star mass 0.1$-$0.3 $M_{\odot}$). 
BW have heavily degenerate companions, being evaporated by the pulsar radiation, and are in the path of becoming isolated MSP, while RB have partially degenerate companion stars and are supposed to be systems in the transition between the accretion-powered and the rotation-powered phase, in which the mass transfer from the companion has temporarily halted 
 \citep[see ][for a review]{roberts13}. A recent, important achievement has been the  discovery of transitional systems, switching between a rotation-powered regime in which they behave as a ``canonical'' RB MSP, and an accretion-powered regime, in which they behave as a peculiar, subluminous low-mass X-ray binary \citep[see ][for a recent review]{papitto22}.
All of these systems are crucial to understand the overall pulsar recycling scenario 
and the ablation process
\citep[see ][for a review]{bhattacharya91}; they are formidable laboratories for studying the acceleration, composition and shock dynamics of the highly relativistic pulsar winds \citep[see e.g. ][]{vandermerwe20}. 

The target of this investigation, PSR J1311$-$3430, was discovered in the $\gamma$-rays ($E>100$ MeV) thanks to a blind search for pulsations in {\em Fermi/LAT} data \citep{pletsch12} and soon after it was also detected as an eclipsing radio pulsar \citep{ray13}. It is an energetic millisecond pulsar with $P=2.5$ ms and spin-down luminosity $\dot{E}_{rot}=5\times10^{34}$ erg s$^{-1}$, in a  tight ($P_B\sim93.8$ min) binary system \citep{pletsch12} with a very low-mass star ($M\sim0.01\,M_{\odot}$). The companion star's rotation is tidally locked at the orbital period; the star side facing the pulsar is heated to $\sim14,000$ K, while the far side is much cooler \citep[$\sim4500$ K, see][]{romani12b}. This, coupled to ellipsoidal variations due to the tidal deformation of the star, produces a very large ($\sim4$ mag) modulation in flux and in colour of the optical counterpart at the orbital period \citep{kataoka12,romani12a}. The pulsar radiation is also powering an intense and highly variable evaporative wind from the companion star, which appears to be fully stripped of hydrogen \citep[H abundance $n(H)<10^{-5}$, see][]{romani12b,romani15}.
Steady X-ray emission is observed from the system in the 0.2$-$10 keV energy range, with a power law spectrum (photon index $\Gamma\sim1.6$) and little or no orbital variability, generally interpreted as produced at the intrabinary shock between the pulsar and the companion star winds \citep{kataoka12,romani12b,an17}.

Dramatic flaring activity  is seen in the optical and near infrared \citep[up to 6 times the flux at the peak of the orbital modulation,][]{romani12a}, originating on the companion star surface and likely powered by the star's magnetic field. Intense flares are also observed in the soft X-rays \citep[up to $\sim10$ times the quiescent level,][]{kataoka12,an17}, possibly as a further manifestation of the release of energy stored in the companion star's magnetic field. Flares in different energy ranges are correlated and may occur at any orbital phase, with a duty cycle in the 10\%-20\% range \citep{an17}. 

We are carrying out a large project aimed at studying the temporal properties of soft X-ray sources \citep[see ][]{deluca21} listed in the X-ray Multi-mirror Mission ({\em XMM-Newton}) serendipitous source catalog. In this context, we discovered a very unusual phenomenon in 
an archival
observation of PSR J1311$-$3430. We describe the peculiar, observed variability pattern in the next section. Section ~\ref{sect:mwl} gives detail on the behaviour of the source in the optical and near ultraviolet band, as derived from simultaneous observations. Possible interpretations and implications of the phenomenon are discussed in Section ~\ref{sect:discussion}.

\section{Peculiar X-ray pulses from PSR J1311$-$3430}
We analized an {\em XMM-Newton} observation of PSR J1311$-$3430 performed on 2018, February 9 (72 ks exposure time). Results from an older observation {\em XMM-Newton} will be given in Sect.~\ref{olderxobs}.  We used data collected from the pn camera \citep{strueder01} and from the two MOS cameras \citep{turner01} of the European Photon Imaging Camera (EPIC) instrument. Using an updated version of the EXTraS software \citep{deluca21}, we generated a background-subtracted light curve of the source in the 0.2$-$10 keV energy range 
using all exposure time -- the EXTraS pipeline features a detailed modelling and subtraction of the time-variable EPIC instrumental background. As shown in Figure ~\ref{mwl} (top panel), PSR J1311$-$3430 undergoes an initial, rapidly decaying trend. After a phase of quiescence with little or no variability, the  behaviour changes and a very peculiar series of six ``pulses'' is seen in the second half of the light curve, the last pulse being truncated at the very end of the observation. 
We may safely exclude an instrumental origin, as the profile of the pulses does not correlate with the variability of the particle background, and no similar pulses are seen in the light curves of other sources in the EPIC field of view.
The regular time spacing between the pulses is apparent, in spite of variability in the peak flux and in the time profile of different pulses.

\subsection{Temporal properties}
\label{sect:timing}

To better characterise the temporal properties of the pulses, we extracted the Power Density Spectrum (PDS) from the light curve of 2018, February after excluding the first portion\footnote{Results do not change if the quiescent phase is also excluded.} (about 10 ks of data), dominated by the bright flare. Fitting a Lorentzian function to the main peak in the resulting PDS yields a central frequency $\nu_c=1.346\pm0.007\times10^{-4}$ Hz, corresponding to a time scale of $\sim7430$ s ($\sim124$ min) -- this is the recurrence time of the pulses. The full width at half maximum (FWHM) of the Lorentzian is $1.8\pm0.2\times10^{-5}$ Hz, which yields a nominal quality factor $Q=\nu_c/\Delta\nu=7.5$, where $\Delta\nu$ is the FWHM of the peak in the PDS. We note that the FWHM of the peak is  consistent with the frequency resolution of the data ($\sim1.7\times10^{-5}$ Hz), so that results on the coherence of the signal should be taken with caution.

 We also performed epoch folding of the same portion of the light curve (the initial flare being excluded), with different trial periods. We fitted a $sinc^2$ function (appropriate for a sinusoidal signal) to the very prominent peak in the resulting periodogram and we evaluated $P=7450\pm50$ s, the error being computed according to \citet{leahy87} and to be taken with 
 caution
 because of the non-sinusoidal shape of the pulses\footnote{The error is also consistent with the uncertainty on the position of the peak as recovered during the $sinc^2$ fitting.}. The coherence of the modulation was investigated by measuring peak time delays along the light curve by correlating a one-period long data segment with a template of the pulse shape produced by folding the light curve at $P=7450$ s. The rms variation relative to a fully coherent modulation is $\sim440$ s, a factor $\sim10$ larger than the  $1\sigma$ statistical uncertainty on the determination of each pulse time delay. 
 Although the data are consistent with a strictly periodic signal, considering the small number of pulses and their variable profile we cannot exclude that the phenomenon is quasi-periodic.

\subsection{Spectral properties}

We studied the energy spectrum of the source and its variability as a function of the time, to search for possible spectral changes associated with the series of pulses (as well as with the initial flare). 

As a first step, we generated energy-resolved light curves in the 0.2$-$1.5 and 1.5$-$10 keV energy ranges (each including about 50\% of the overall source photons) and we computed the hardness ratio as a function of the time. This turned out to be fully consistent with a constant (p value $>95\%$), suggesting that no significant spectral changes occur in spite of the peculiar, large flux variability. We also folded the hardness ratio curve at P=7450 s, but no hints of modulation were found.

Then, we performed time-resolved spectroscopy. Source photons were selected from a circle with a radius of $25''$; background photons were extracted from a source-free region located in the same CCD as the source. Ad-hoc response and effective area files were generated with the SAS software\footnote{\url{https://www.cosmos.esa.int/web/xmm-newton/what-is-sas}}. Spectral modelling was performed with Xspec\footnote{\url{https://heasarc.gsfc.nasa.gov/xanadu/xspec/}}. We quote uncertainties at the 68\% confidence level for a single parameter of interest.
First, we studied the time-averaged spectrum, which turned out to be well described (p value=$2.2\times10^{-2}$) by a
power law with photon index $\Gamma=1.53\pm0.04$, absorbed by a column $N_H=(5\pm1)\times10^{20}$ cm$^{-2}$. The observed flux in the 0.2$-$10 keV energy range is $(1.9\pm0.1)\times10^{-13}$ erg cm$^{-2}$ s$^{-1}$.
Then, we selected time intervals corresponding to the initial flare, to the peaks of individual pulses and to the low-flux, quiescent level, based on a simple count rate threshold, and we extracted three spectra. No significant spectral
changes could be detected. We show in Table~\ref{table1} results from a simultaneous fit to the three spectra, linking the column density to be the same and leaving the photon index and normalization as free parameters (p value=0.83). The emission is always consistent with the time-averaged spectrum.

\vspace{3mm}

\begin{table}[hbt!]
\centering       
\caption{Results of {\em XMM-Newton} time-resolved spectroscopy for the time intervals of the initial flare, the peaks of pulses (combined) and the low-flux, quiescent level. The three spectra are simultaneously fitted linking the $N_H$ value. 
The time-averaged results are also shown for comparison.
}
\begin{tabular}{cccc}
\hline  
\hline
Interval & Exp.(pn/MOS) & $N_H$ & $\Gamma$\\
 & ks & $10^{20}$ cm$^{-2}$ & \\
\hline
Flare & 0.5/1.9 & 5$\pm$1 (linked) & 1.60$\pm$0.13\\
Pulses & 13.1/15.1 & 5$\pm$1 (linked) & 1.54$\pm$0.07\\
Quiescence & 39.3/44.7 & 5$\pm$1 (linked) & 1.55$\pm$0.07\\
\hline
Total & 60.7/70.1 & 5$\pm$1 & 1.53$\pm$0.04\\
\hline
\hline                  
\end{tabular}\label{table1} 
\end{table}

\subsection{Older, archival X-ray observations}
\label{olderxobs}
We also studied the only  previous {\em XMM-Newton} observation of the system, performed on 2014, August 2 (120 ks). The source background-subtracted light curve features a rather brigth flare at the beginning of the observation, but does not show any hint of recurring pulses with a regular time spacing \citep[Figure 5b of][shows the source light curve from that dataset]{an17}.
We repeated both the {\em PDS} analysis and the folding analysis on that light curve. No significant periodic or quasi-periodic signals were detected in either method. 

For completeness, we also performed spectroscopy of the 2014, August observation, following the same procedure described above. In this case, the spectra of the flaring period ($\sim$15 ks) and of the quiescent period ($\sim$100 ks) turned out to be different. 
A simultaneous fit where both the normalizations and the photon indexes were allowed to vary between the two intervals results in an acceptable fit (p value=$10^{-3}$) with $N_H=(2\pm2)\times10^{20}$ cm$^{-2}$, $\Gamma_{Flare}=1.23\pm0.05$ and $\Gamma_{Quiesc}=1.57\pm0.07$. Our results are fully consistent with those of \citet{an17}.
We note that the flaring period caught by this observation is remarkably longer ($\sim$10 times), although fainter, than the one seen in the 2018 observation and this results in smaller errors on the best fit parameters. The photon index of the two flaring periods is compatible at the $2\sigma$ level.

Other X-ray observations of PSR J1311$-$3430 were performed with {\em Chandra}, {\em Suzaku} and {\em Swift}/XRT, but they lack the sensitivity and/or the uninterrupted exposure time needed to firmly identify similar X-ray pulses \citep[see ][]{kataoka12,arumugasamy15,an17}.

\section{Optical and near Ultraviolet behaviour}
\label{sect:mwl}
We investigated the multiwavelength properties of the pulses, taking advantage of available data collected simultaneously with the {\em XMM-Newton} observation of 2018.

\begin{figure*}
\begin{center}
\includegraphics[angle=-90,width=17cm]{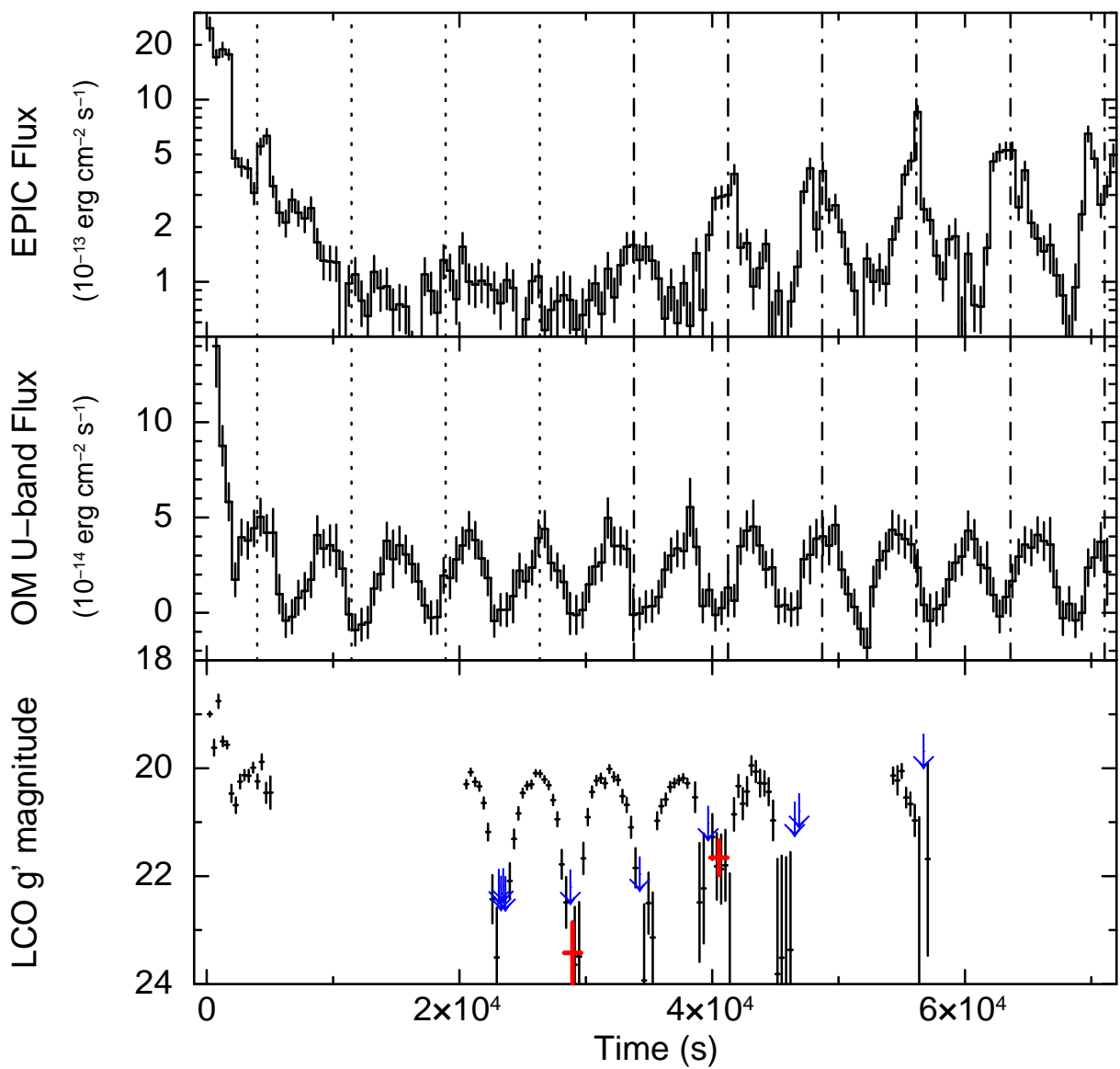} 
\end{center}
\caption{Multiwavelength light curves of PSR J1311$-$3430. Top panel: 0.15-10 keV energy range, {\em XMM-Newton}/EPIC data (background-subtracted, 500 s binning). Count rate to flux conversion was performed using results of time-averaged spectroscopy. The approximate time of the centers of the six ``pulses'' is marked by vertical dashed-dotted lines (adopting 7450 s as the recurrence time -- see Sect.~\ref{sect:timing}).  We also extended to earlier times the expected time of arrivals (dotted lines) to show the  possible alignement of a further peak at T=4000 s with the pulses. The first pulse peak occurs at orbital phase 0.08+/-0.03 (as computed from the U-band light curve, phase 0.75 marking the companion star superior conjunction). Middle panel: U band, {\em XMM-Newton}/OM data (background-subtracted, 500 s binning). Bottom panel: g' band, Las Cumbres Observatory - 1m telescope data; black points represent 103 observations lasting 300 s; red points correspond to the sum of 4 observations (totalling 1200 s integration) collected around the times of two different g' light curve minima, the first associated with quiescent X-ray emission (g'$=23.4\pm0.5$) and the second associated with an X-ray pulse (g'$=21.6\pm0.3$); upper limits are shown as blue arrows. In all panels time is measured in seconds since 2018-02-09 22:18:50.8. Error bars show uncertainties at the $1\sigma$ confidence level.} 
\label{mwl}
\end{figure*}

\subsection{Near-ultraviolet observations with XMM/OM}
\label{om}
During the {\em XMM-Newton} observation of 2018, February 9, 
the Optical Monitor observed PSR J1311$-$3430 in Fast Mode with the U filter ($\lambda=344$ nm, $\Delta\lambda=84$nm). 
We performed a standard data reduction and analysis using the dedicated SAS software. Count rate to flux conversion was performed using the conversion factors derived from observations of white dwarf standard stars provided by the calibration team\footnote{\url{https://www.cosmos.esa.int/web/xmm-newton/sas-watchout-uvflux}}.

The resulting light curve (see Figure~\ref{mwl}, middle panel) shows the well-known modulation at the orbital period of the system, with maxima corresponding to the companion star superior conjunction (i.e. its heated side seen face-on). The only apparent deviation from this trend is an obvious  counterpart of the X-ray flare occurring at the beginning of the observation  (the recovery from the flare having a different time profile in the two energy ranges). 

We searched for possible UV variability correlated to the X-ray pulses. We selected time intervals corresponding to the peaks of the X-ray pulses and to the X-ray quiescent level in the X-ray light curve shown in Fig.~\ref{mwl}: for the pulses, we adopted a flux threshold $F_X>3\times10^{-13}$ erg cm$^{-2}$ s$^{-1}$ as a simple criterion, which yielded $16\times500$ s bins in the last 5 pulses; for the quiescence, we selected the portion of the light curve between Time=10,000 s and Time=30,000 s. We modelled the UV light curve in the two time intervals as the sum of a constant and of a sin function. A simultaneous fit forcing all parameters to be the same yields a very good description of the data (p-value$>$0.5). Although this suggests that there are no significant variations in the UV emission correlated to the pulses, we also allowed the constant term $C$ to vary and we computed $\Delta F_{UV}/F_{UV,quiesc}$ using the best fit values for $C_{pulses}$ and $C_{quiesc}$, as $(C_{pulses}-C_{quiesc})/C_{quiesc}=0.32\pm0.19$. For a comparison, the variation of the X-ray flux in the two selected intervals is $\Delta F_X/F_{X,quiesc}=5.00\pm0.35$.
As a further exercise, we removed the orbital modulation by dividing the OM lightcurve (binned at $P_B/16 = 351.6$ s) by a template obtained from the data folded at the orbital period in 16 bins. The resulting detrended light curve was then folded at the periodicity of the X-ray pulses. No significant deviations from a constant profile were found.

\subsection{Optical observations with LCO}
\label{lco}
Ground-based observations of PSR J1311$-$3430 were performed\footnote{proposal ID: NOAO2018A-021} simultaneously with the {\em XMM-Newton}, 2018 February 9 observation, with the Sinistro camera at the 1m telescope at Las Cumbres Observatory (LCO\footnote{\url{https://lco.global/}}) in the g' band ($\lambda=477$ nm, $\Delta\lambda=150$ nm). 103 observations with 300 s exposure time were performed.  We retrieved images reduced with the BANZAI pipeline\footnote{\url{https://lco.global/documentation/data/BANZAIpipeline/}} from the LCO data archive\footnote{\url{https://archive.lco.global/}}. We ran a standard source detection using the sextractor software \citep{bertin96}. The target was detected in part of the observations only, thus we also performed forced aperture photometry at its position. With the main aim of investigating the flux variability as a function of the time, we performed a quick photometric calibration using a set of USNO-B1 stars. 

The resulting  light curve is shown in Figure~\ref{mwl} (bottom panel). The orbital modulation is clearly visible, minima being poorly constrained because of a low signal-to-noise. The recovery from the initial bright flare is also apparent. Unfortunately, there is only partial LCO coverage of the time intervals where the X-ray pulses are seen, because of gaps in the series of observations and/or of poor atmospheric conditions. To search for variability correlated to the X-ray pulses, we combined data from multiple observations corresponding to two different optical minima, one aligned with an X-ray pulse and the other simultaneous with X-ray quiescent emission and repeated the analysis. Although the signal to noise is low, we find some evidence of enhanced g' flux associated with the X-ray pulse (g'$=21.6\pm0.3$) with respect to the X-ray quiescent level (g'$=23.4\pm0.5$). The difference between the two fluxes  corresponds to a $50\%\pm20\%$ variation with respect to the average value seen during the orbital modulation. For a comparison, the  X-ray fluxes in the two selected intervals differ by a factor $4.5\pm0.8$). 

\subsection{The initial flare}
\label{mwlflare}
We also studied the multiwavelength properties of the initial flare (see Figure~\ref{mwl}). In the first 500 s of the OM observation, the U band flux is a factor $7\pm1$ higher than the average level measured when the orbital modulation is seen. In the same time interval, the g' light curve displays a factor $5.7\pm0.3$ increase with respect to the average level (measured across the orbital modulation, as in the case of the U band). In the X-rays, the flux in the same time interval is a factor $24\pm3$ higher than the quiescent level. In all cases, the flux variation is computed as ${(F_{peak}-F_0)}/F_0$ where $F_0$ is the average or quiescent level. Results for the initial flare and for the pulses are compared in Table~\ref{table:mwl}. These suggest a different spectral energy distribution for the two phenomena, the pulses having higher $F_X/F_U$ and $F_X/F_{g'}$ ratios than the flare.

\begin{table}[hbt!]
\centering       
\caption{Multiwavelength properties of the pulses and of the initial flare. For each energy range, we computed the relative flux variation for the pulses as ${(F_{peak}-F_0)}/F_0$, where $F_{peak}$ is the flux at the peak of the pulses and $F_0$ is the quiescent flux level (see Sect.~\ref{om} and ~\ref{lco}). The same relative flux variation was computed for the initial flare (see Sect.~\ref{mwlflare}).
}
\begin{tabular}{ccc}
\hline  
\hline
Energy band & Pulses & Flare \\
& $\Delta~F/F_0$ & $\Delta~F/F_0$ \\
\hline
X-rays & $5.00\pm0.35$ & $24\pm3$ \\
U band & $0.32\pm0.19$ & $7\pm1$ \\
g' band & $0.5\pm0.2$ & $5.7\pm0.3$ \\
\hline
\hline                  
\end{tabular}\label{table:mwl} 
\end{table}

\section{Discussion}
\label{sect:discussion}
Interpretation of the periodic X-ray pulses is challenging. 
First, their recurrence time of $\sim124$ min is puzzling and requires the identification of a characteristic frequency in the system different from the $\sim94$ min orbital periodicity of the binary. 
Second, the phenomenon is transient in nature. The series of pulses starts in the middle of the 2018 {\em XMM-Newton} observation, after a phase of quiescent X-ray emission lasting several hours. It is not seen in the longer {\em XMM-Newton} observation performed in 2014.
Third, the pulses are seen in very different geometric configurations of the binary system, occurring at either the companion star superior conjunction, inferior conjunction, ascending node or descending node. 
Fourth, the energetics of the pulses is very large. Assuming a distance of 1.4 kpc \citep{ray13} yields a peak luminosity of order $\sim3.3\times 10^{32}$ erg s$^{-1}$ (close to 1\% of the pulsar $\dot{E}_{rot}$) and an integrated energy of several $10^{35}$ erg per pulse. 
Fifth, in spite of a very large variation in X-ray flux, 
the energy spectrum does not change when the pulses are seen -- emission being always well described by a flat power law with a photon index of $\Gamma\sim1.6$.

PSR J1311$-$3430 was already known to display large flares (up to $10^{33}$ erg s$^{-1}$) in the optical and X-ray energy ranges \citep{kataoka12,romani12a}.  
This behaviour was also seen at the beginning of the 2018 XMM-Newton observation (see Fig.~\ref{mwl}). As a first possibility, the pulses could be a (quasi)-periodic series of such flares. 
It is difficult to compare the two phenomena because the energy distribution of flares is not precisely known.
 \citet{an17} showed that the flaring emission in the X-rays and in the optical/near ultraviolet is correlated (indeed, as seen for the initial flare in Fig.~\ref{mwl}). A large flare was caught by \citet{romani15} during optical spectroscopy -- in that case, the optical spectrum was well described as a (transient) atmosphere of a He-dominated star with a temperature of $\sim39,000$ K,
 radiated from 20\% of the surface of the companion star. This suggests that emission was powered by energy released below the photosphere, likely related to the magnetic field of the companion star \citep{romani15,an17}. It is unclear whether this picture may describe all flares from the source. We note that similar flares at X-ray and optical wavelengths are observed in other similar systems \citep[see e.g. ][and references therein]{halpern22} -- flaring emission in the candidate redback 1FGL J0523.5$-$2529 has a non-thermal shape in the optical range, suggesting that flares come from above the photosphere of the star, possibly related to reconnection of striped pulsar magnetic field compressed in the intrabinary shock \citep{halpern22}. 
In any case, the possible analogy of pulses and flares seems hardly reconcilable with the lack of any flux variation in the U and in the g' band\footnote{The possible enhanced g' emission associated to the pulses could be related to He I line emission from the wind, as observed by \citet{romani15}.}, 
correlated with the X-ray pulses (see Table~\ref{table:mwl}). Some X-ray spectral variation associated to the pulses would also be expected (indeed, as observed for the already known X-ray flares from the system). Although the small number of observed pulses and flares prevents one to draw firm conclusions, current multiwavelength data suggest that pulses and flares are two different phenomena. This is also supported by the possible evidence for  episodes of bright He line emission having the same $\sim2$ hr recurrence time of the pulses, seen in non-simultaneous optical spectroscopy (see last paragraph).

We also consider the possible analogy of PSR J1311$-$3430 pulses with flaring behaviour observed in  other binary millisecond pulsars.
Transitional RB systems show peculiar X-ray moding during their LMXB (accreting) phase, one of the modes consisting in the emission of series of erratic flares \citep{papitto22}. Such flares are seen to occur on a variety of time scales, from few s to $\sim1$hr, and occasionally display a quasi-periodic pattern, somehow reminiscent of the series of pulses seen from PSR J1311$-$3430 \citep[see e.g.][]{jaodand21,strader21}. As a matter of fact, the physics of the flaring mode in transitional systems is not understood. Are the X-ray pulses from PSR J1311$-$3430  powered by the same (unknown) mechanism? Multiwavelength phenomenology does not support this picture. Transitional pulsars in their flaring mode show correlated emission in the X-rays and in the near infrared to near ultraviolet range, which is not seen in the case of our pulses (see Sect.~\ref{sect:mwl}). The luminosity of X-ray flares from transitional pulsars \citep[$2-7\times10^{34}$ erg s$^{-1}$, ][]{papitto22} is also higher by $\sim2$ orders of magnitude than that of the pulses of PSR J1311$-$3430. 
Should this analogy be proven by future observations, it would have extremely interesting implications
since a common mechanism would be at work in very different contexts.  In fact the flaring mode in transitional MSP is seen when they are accreting, while accretion is not occurring in PSR J1311$-$3430, where the momentum flux of the pulsar overwhelms that of the companion's wind.

We may also explore other possible explanations for the pulses within the standard picture for BW systems.
The lack of  X-ray spectral changes across the complex variability pattern suggests that pulses could be produced by the same process generating the ``quiescent'' emission: synchrotron emission at the intrabinary shock (IBS) between the pulsar's relativistic wind and the companion star's massive wind. 
The position and shape of the IBS depend on the momentum balance between the pulsar and companion winds, as well as on the orbital velocity \citep{arons93}. These quantities also determine the observed high-energy flux from the IBS, depending on the fraction of the  pulsar spin-down power intercepted by the IBS (setting the IBS luminosity) and on Doppler boosting of the X-ray emission when the mildly relativistic post-shock plasma flow is directed along the line of sight \citep[see e.g. ][]{arons93,kandel19,vandermerwe20,cortes22}. 
Interpretation of the pulses within this scenario is not easy. Large inhomogeneities in the companion star wind would be required to trigger major changes in the IBS luminosity and/or in its emission pattern. Explaining the periodicity would be more difficult: on the one side, an unknown (transient) mechanism producing periodic variations in the companion star wind properties would be required; on the other side, it seems hardly conceivable to observe pulses with similar properties associated with totally different orbital phases of the system, in view of the major role of the geometry of sight in shaping the IBS observed phenomenology. 

To explain the regular recurrence time of the phenomenon, we could invoke the existence of a third body in the system \citep[e.g. a planet, as in the case of PSR B1257$-$20,][]{wolszczan92}, interacting with the pulsar radiation and wind and producing (e.g. through a shock/via non-thermal bremsstrahlung) the observed X-ray pulses. With an orbital period of $\sim124$ min, it would stay on an orbit only $\sim20\%$ larger than the one of the known companion star, which would pose severe problems in explaining the system's long-term stability, the lack for any direct evidence of the putative third body in the near infrared/optical and in the timing phenomenology of the millisecond gamma-ray pulsar\footnote{Possible evidence for a very low-mass third body gravitationally tied to the Black-widow Pulsar PSR J1555$-$2908 based on accurate $\gamma-$ray timing was recently reported by \citet{nieder22}.}, as well as the transient behaviour of the pulses. The possibility of an object sitting in a larger orbit ($P\gg124$ min) could also be considered; in that case, one would need a mechanism ``illuminating'' the object once per 124 min cycle, which seems a rather contrived picture. We also note that searches for planetary companions to millisecond pulsars, taking advantage of the very accurate timing achievable for such systems, proved planets to be a rarity \citep[see e.g.][]{behrens20}.

As a further alternative, we could consider an orbiting, dense ``blob'' of ablated gas. Such a blob, originating from inhomogeneities in the companion star wind and/or from instabilities in the post-shock flow, should be confined within the region occupied by the shocked companion star wind -- expected to have the shape of the apex of an Archimedean spiral, bent by Coriolis force and radiation pressure \citep{rasio89,arons93} -- and would participate in the quiet outward motion of the shocked wind at a speed comparable to the star orbital velocity \citep{romani15}. 
Indeed, BW and RB systems show both regular eclipses as well as ``mini-eclipses'' that occur at sporadic orbital phases, are not stable in orbital phase from one orbit to the next, and likely indicate such blobs of intra-binary material \citep[see e.g. Fig. 5 of ][]{archibald13}. 
However, it would be difficult to explain the production of copious X-rays in such a slow flow, and the persistence of a regular recurrence time in the emission over $\sim8$ system revolutions.   

To assess the actual viability of the above hypothesis, detailed calculations and modelling would be needed, which are beyond the scope of this letter.

An important piece of information would be the evaluation of the duty cycle of the X-ray pulses. In the soft X-rays, the only available observations with the sensitivity and long, uninterrupted exposure needed to firmly identify the phenomenon are the two {\em XMM-Newton} ones. At other wavelengths, {\em Fermi/LAT} data do not have the sensitivity to detect individual pulses. Turning to lower energies,
we note that \citet{romani15} published  trailed spectra to show the spectral evolution of PSR J1311$-$3430 in the optical range as a function of the orbital phase, based on Keck/LRIS data, collected in 2013, February and covering $\sim3$ orbital periods (see their Figure 2). A very interesting feature is a series of three bright ``flares'',  consisting of He I line emission only (thus, coming from circumstellar gas), clearly apparent in the red branch of the spectrum. 
It can be easily seen that their time spacing is about the same that we observe between our X-ray pulses. Thus, it is very tempting to relate these three He I line emission ``flares'' to the same phenomenon we observe in {\em XMM-Newton} 2018 data -- He I line emission being possibly excited by  X-ray periodic pulses. Under this assumption, on the one side study and modelling of the Keck 2013 data could constrain the position and velocity of the gas illuminated by the X-ray pulses, possibly constraining the emitting region of the pulses themselves. On the other side, the periodic mechanism triggering the pulses would have a larger duty cycle and thus there could be important chances to re-observe it. Simultaneous X-ray observations and optical spectroscopy could confirm the picture and shed light on this puzzling phenomenon, potentially very relevant for our understanding of BW systems and of the overall evolution of MSP.

\begin{acknowledgements}
This work is based on observations obtained with {\em XMM-Newton}, an ESA science mission with instruments and contributions directly funded by ESA Member States and NASA. This work makes use of observations from the Las Cumbres Observatory global telescope network, performed with the Sinistro camera at the 1m Telescope. We thank an anonymous referee for his/her helpful comments.  We acknowledge financial support from ASI under ASI/INAF
agreement N.2017-14.H.0. We also acknowledge  support via ASI/INAF Agreement n. 2019-35-HH and PRIN-MIUR 2017 UnIAM (Unifying Isolated and Accreting Magnetars, PI S.~Mereghetti).
\end{acknowledgements}

\bibliographystyle{aa}
\bibliography{j1311}

\end{document}